\documentclass[floatfix,showpacs,amsmath,amssymb,letterpaper,groupaddresses,superscriptaddress]{article}
\usepackage{graphicx}
\usepackage{latexsym}
\usepackage{amsmath}
\usepackage{amssymb}
\usepackage{epsfig}
\usepackage{graphicx}
\usepackage{graphicx}
\usepackage{amssymb,amsmath,times}
\usepackage{hyperref}
\usepackage{color}

\def\be{\begin{equation}}
\def\ee{\end{equation}}
\def\ba{\begin{eqnarray}}
\def\ea{\end{eqnarray}}
\def\la{\langle}
\def\ra{\rangle}

\usepackage{epsfig}
\usepackage{graphicx}
\usepackage{multirow}
\usepackage{caption}
\begin{document}
\begin{center}
{\Large \bf A quantum phase transition from $Z_2 \times Z_2$  to $Z_2$ topological order}\\
\vspace{1cm}Mohammad Hossein Zarei\footnote{email:mzarei92@shirazu.ac.ir}\\
\vspace{5mm}
\vspace{1cm}Physics Department, College of Sciences, Shiraz University, Shiraz 71454, Iran
\end{center}
\vskip 3cm
\begin{abstract}
Although the topological order is known as a quantum order in quantum many-body systems, it seems that there is not a one-to-one correspondence between topological phases and quantum phases. As a well-known example, it has been shown that all one-dimensional (1D) quantum phases are topologically trivial\cite{spt}. By such facts, it seems a challenging task to understand when a quantum phase transition between different topological models necessarily reveals  different topological classes of them. In this paper, we make an attempt to consider this problem by studying a phase transition between two different quantum phases which have a universal topological phase. We define a Hamiltonian as interpolation of the toric code model with $Z_2$ topological order and the color code model with $Z_2 \times Z_2$ topological order on a hexagonal lattice. We show such a model is exactly mapped to many copies of 1D quantum Ising model in transverse field by rewriting the Hamiltonian in a new complete basis. Consequently, we show that the universal topological phase of the color code model and the toric code model reflects in the 1D nature of the phase transition. We also consider the expectation value of Wilson loops by a perturbative calculation and show that behavior of the Wilson loop captures the non-topological nature of the quantum phase transition. The result on the point of phase transition also show that the color code model is strongly robust against the toric code model.
\end{abstract}
\section{Introduction}
After the symmetry breaking theory of Landau \cite{sym}, many thought science has reached a comprehensive understanding of different phases of matter. However, a new concept emerged in condense matter physics by quantum Hall effect \cite{hall}, quantum spin fluids \cite{spin1, spin2, spin3, 23} and superconductors \cite{20, 21, 22} which was called topological order. Topological order is a different phase of matter which can not understood by Landau theory so that one can find different topological phases with the same symmetries \cite{wen}. \\
The ground state of a quantum system with topological order exhibits a long-range entanglement \cite{long}. Specifically, the topological entanglement entropy has been defined as a non-local order parameter of topological order \cite{16, 17} so that it can characterize a quantum phase transition from a topological phase \cite{18}. Another interesting property of a topological phase is the behavior of the Wilson loop where the expectation value of the Wilson loop is related to perimeter of the loop in a topological phase, while it is related to area of the loop in a non-topological phase \cite{18}.\\
Long-range entanglement of the ground state of a topological matter also leads to some interesting and exotic properties which have not been seen in symmetry breaking phases. Robust degeneracy of the ground state and exotic statistics of excitations of the system, which are called anyons, are two important characteristics of the topological phases \cite{wen2, wen3, wen4}. Braiding anyons leads to an arbitrary phase factor on wave function of the system in the abelian models and a unitary operator in the non-abelian models \cite{g, h}. The robustness of degeneracy of the ground state and braiding operators against local perturbations is a key property of the topological phases which has converted them to important candidates for fault-tolerant quantum computation \cite{pr, fr, kitaev}.\\
In spite of many improvements on defining different characteristics of a topological phase, recognizing different topological classes in different quantum systems \cite{a, b, c, d, e, f} is still an open problem which has received much deal of interest. Since topological order is a kind of quantum order, it is clear that the same quantum phases have also the same topological phases. Consequently, considering quantum phase transition between different topological models is a useful approach for classification of topological models. Specifically, there are some recent papers which show a quantum phase transition can well reveal different topological properties of two different topological phases \cite{qq, karimipour}. \\
This thought that a quantum phase transition reveals necessarily different topological classes of the quantum phases is challenged specifically by 1D quantum models. It has been shown that all 1D quantum phases belong to a trivial topological phase \cite{spt} so that one can find a quantum phase transition between two different 1D quantum models which are topologically trivial \cite{pasc, va}. In fact, Since topological signature of topological phases such as topological entanglement entropy or expectation value of the Wilson loop operators can not be defined for a 1D model, topological order of 1D models only defines as a symmetry protected topological phase \cite{spt}.\\
Moreover 1D quantum models which are in the same topological phases, there are also two-dimensional (2D) quantum models which belong to a universal topological phase. A well-known example of universal topological phases has been seen in the toric code model (TC) with $Z_2$ topological order \cite{kitaev} and the color code model (CC) with $Z_2 \times Z_2$ topological order \cite{bombin}. Since the TC and CC model have different gauge symmetries, degeneracies of the ground state of them are different. While the TC model provides a four-fold degenerate ground state which can be used as a robust quantum memory, degeneracy of the ground state of the CC model is sixteen-fold. Furthermore, unlike the TC model, there is also possibility of applying unitary Clifford group in a topological way \cite{bombin}. Recently, there is also an experimental realization of the color code on trapped-ion qubits \cite{ex}.\\
Although different degeneracies of the CC model and the TC model show that they are in two different quantum phases, other topological characteristics of both models are completely the same. Especially, it is possible to show a CC model and two copies of the TC model are in the same quantum phases \cite{fold, fold1}. In \cite{fold}, the authors have emphasized that since there is an adiabatic evolution without quantum phase transition between a CC model and two copies of the TC model, they have the same topological phases. This result shows that different quantum phases of the TC and the CC model are completely related to their different gauge symmetries and has not a topological nature.\\
By above facts, there is a good chance that one considers how the same topological characteristics of the CC and the TC model reflects in the properties of a quantum phase transition between them. To this end, in this paper we consider a CC model on a hexagonal lattice besides a single TC model instead of two copies of it on the same lattice. We show the universal topological phase of the CC and the TC model reflects in properties of the quantum phase transition where we show that the quantum phase transition has a 1D nature. Our result is also another proof for the universal topological phase of the CC model and the TC model. We emphasize that our proof has a recent and important point where generally shows how the universal topological phase of two 2D topological models can be revealed even in presence of a quantum phase transition. \\
To deriving results, we define a specific version of the TC model on the hexagonal lattice. We show that a model Hamiltonian as interpolation of the CC and the TC model is mapped to many copies of 1D Ising models in transverse field which belong to the 2D classical Ising universality class. The mapping is based on re-writing the Hamiltonian of the model in a new complete basis. Our method is equivalent with unitary transformation on the Hamiltonian which does not change spectrum of the model. We also study the behavior of Wilson loops and explicitly show that the expectation value of the Wilson loops in our model captures the non-topological nature of the phase transition.\\
 From another point of view, we find the point of the phase transition of our model which is the same as 1D Ising model in transverse field at $\frac{g_t}{g_c}=1$ \cite{one} where $g_t$ and $g_c$ are the couplings of the TC and the CC models, respectively. It shows when we add the TC Hamiltonian as a small perturbation against the CC model, such a perturbation can not ever change the quantum phase of the CC model. The quantum phase transition occurs only when a considerable perturbation $g_t >g_c$ is applied. Such a result show that unlike the small robustness of the CC model against local perturbations such as magnetic field or Ising interaction \cite{i, zarei, kargarian, lan, jah}, it is strongly robust against a topological perturbation like the TC model.\\
In section (\ref{se1}), we review different properties of the topological phases of the TC and the CC model. In Section (\ref{se3}), we present our model as a interpolation of the CC and the TC model. We define both topological models on a hexagonal lattice and qualitatively show how a quantum phase transition happens. In section (\ref{se4}), we use a mathematical method to map our model to 1D Ising models in transverse field which have a well-known phase transition point. Finally, in section (\ref{se5}) we consider the behavior of Wilson loop operators by a perturbative approach to show the quantum phase transition in our model has not a topological nature.
\section{Brief review of the toric code and the color code model}\label{se1}
In this section we review the topological structure of the TC and the CC models. We specifically emphasize on different topological degeneracies and different topological structures of them by representation of the ground states as a loop-condensate states. For more details, there is also a comparative study of these models which has done in \cite{comp}.\\
\subsection{Toric code model}\label{se11}
 A TC model is ordinarily defined as the ground state of a model Hamiltonian on an oriented graph where qubits live on edges of the graph. Two commutative operators $B_p$ and $A_s$ are defined corresponding to each plaquette and vertex of the graph in the following form, See figure (\ref{toric}, left) for a square lattice:
\begin{figure}[t]
\centering
\includegraphics[width=10cm,height=4.5cm,angle=0]{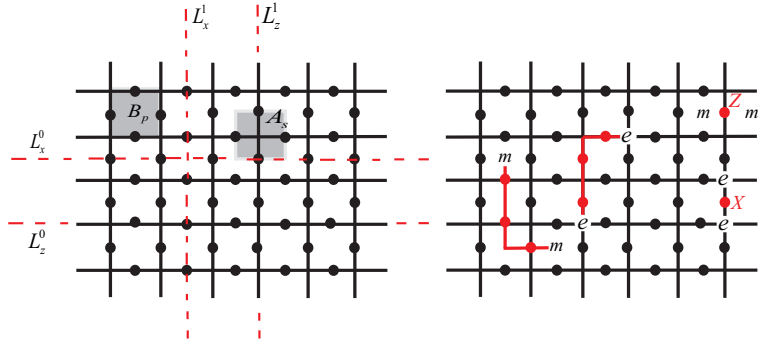}
\caption{(Color online) Left: the plaquette and vertex operators have been shown by two different colors corresponding to vertices and plaquettes of the lattice. There are four non-trivial loop operators which describe topological degeneracy. Right: two charge (flux) anyons are generated by applying $Z$ ($X$) operator on each qubit. By applying a sequence of $Z$ ($X$) operators on qubits it is possible to move a charge (flux) anyon on the lattice.} \label{toric}
\end{figure}
\begin{equation}
B_p =\prod_{i\in p}Z_i~~~,~~~A_s=\prod_{i\in s} X_i
\end{equation}
where $i\in p$ refers to qubits around of a plaquette and $i\in s$ refers to qubits around of a vertex and $Z~,X$ are the Pauli operators. The Hamiltonian of the model is as summation of these operators on all plaquettes and vertices of the graph as the following form:
\begin{equation}\label{e1}
H=-\sum_{p}B_p -\sum_{s}A_s .
\end{equation}
 By the fact that the vertex and plaquette operators commute with each other, it is simple to show the following state is a ground state of Hamiltonian (\ref{e1}):
 \begin{equation}\label{e2}
 |\phi_t \rangle = \prod_{s} (1+B_p)|+++...+\rangle
 \end{equation}
 where the state $|+\rangle$ is eigenstate of the operator $X$ corresponding to eigenvalue $+1$ and we ignore the normalization factor for this state.\\
 There is also a simple representation for the state (\ref{e2}) which helps for better understanding of the topological order of such a state. To this end, let us span the product of operators $(1+B_p)$ on all plaquettes in the relation (\ref{e2}) as the following:
 \begin{equation}\label{e3}
 \prod_{p}(1+B_p)= 1+\sum_{p}B_p +\sum_{p,p'} B_p B_{p'}+...
 \end{equation}
 where the right hand of this relation is the summation of all possible products of the plaquette operators. Since each plaquette operator can be interpreted as a loop operator on the lattice, right hand of the relation (\ref{e3}) is the summation of all possible loop operators. \\Finally, since $Z|+\ra=|-\ra$ where the state $|-\rangle$ is the eigenstate of the operator $X$ corresponding to eigenvalue $-1$, the ground state of the TC model (\ref{e2}) can be interpreted as uniform superposition of all loop constructions of qubits $|-\ra$ in the sea of qubits $|+\ra$ which is called the loop condensation. \\Such a state has a topological order which leads to a robust degeneracy in the ground state of the model when we define the model Hamiltonian on a torus. In fact, on a torus topology, there are also non-contractible loop operators in the form of:
 \begin{equation}
 L_{Z}^{\sigma}=\prod_{i\in L^{\sigma}}Z_i
 \end{equation}
 where $i\in L^{\sigma}$ refers to the qubits which live on a non-contractible loop around the torus in two different directions $\sigma=0~or~1$, see figure (\ref{toric}, left). Therefore, four the following quantum states are the degenerate ground states of Hamiltonian (\ref{e1}):
 \begin{equation}\label{e4}
 |\psi_{i,j}\ra=( L_{z}^{0})^i ( L_{z}^{1})^j |\phi_t\rangle
 \end{equation}
 where the indices $i,j=\{0,1\} $ refer to four different quantum states.\\
 Topological order of the TC model are understood by three important properties, robust degeneracy, non-local order parameter and anyonic excitations.\\ The robust degeneracy is due to this fact that four degenerate ground states (\ref{e4}) can not be converted to each other by any local parameter so the degeneracy is robust against each local perturbation.\\
Another property is that any local operator can not distinguish four degenerate ground states. In fact, there are two non-local operators which have different expectation values in different degenerate ground states. Such operators are defined in the following form:
 \begin{equation}
 L_{X}^{\sigma}=\prod_{i\in L^{\sigma}}X_i
 \end{equation}
 where $i\in L^{\sigma}$ refers to the qubits which live on a non-contractible loop around the torus in two different directions $\sigma=0~or~1$, see figure (\ref{toric}, left). The expectation values of these operators in each one of the four degenerate ground states are as follows:
$$ \la \psi_{00}|L_{x}^{0}|\psi_{00}\ra =1~~,~~\la \psi_{00}|L_{x}^{1}|\psi_{00}\ra =1$$
$$ \la \psi_{01}|L_{x}^{0}|\psi_{01}\ra =-1~~,~~\la \psi_{01}|L_{x}^{1}|\psi_{01}\ra =1$$
$$ \la \psi_{10}|L_{x}^{0}|\psi_{10}\ra =1~~,~~\la \psi_{10}|L_{x}^{1}|\psi_{10}\ra =-1$$
  \begin{equation}
   \la \psi_{11}|L_{x}^{0}|\psi_{11}\ra =-1~~,~~\la \psi_{11}|L_{x}^{1}|\psi_{11}\ra =-1
 \end{equation}
 Therefore these two non-local operators can distinguish the different ground states.\\
  Finally, excitations of the TC model are quasi-particles with anyonic statistics. An excitation is generated by applying the Pauli operators $X$ or $Z$ on a qubit of the lattice, see figure(\ref{toric}, right). An operator $X$ on a qubit does not commute with two plaquette operators which are shared in that qubit and it is interpreted as two flux anyons $m$ in two corresponded plaquettes. Also an operator $Z$ does not commute with two neighbor vertex operators and it is interpreted as two charge anyons $e$ in two corresponded vertices. By applying a string of $X$ ($Z$) operators, a flux (charge) anyon moves on plaquettes (vertices) of the lattice, see figure (\ref{toric}, right). It is simple to show that if a charge anyon winds around a flux anyon, it leads to a minus sign on wave function. Such a factor shows that charge and flux anyons are not  fermions or bosons.
 \subsection{Color code model}
\begin{figure}[t]
\centering
\includegraphics[width=10cm,height=6cm,angle=0]{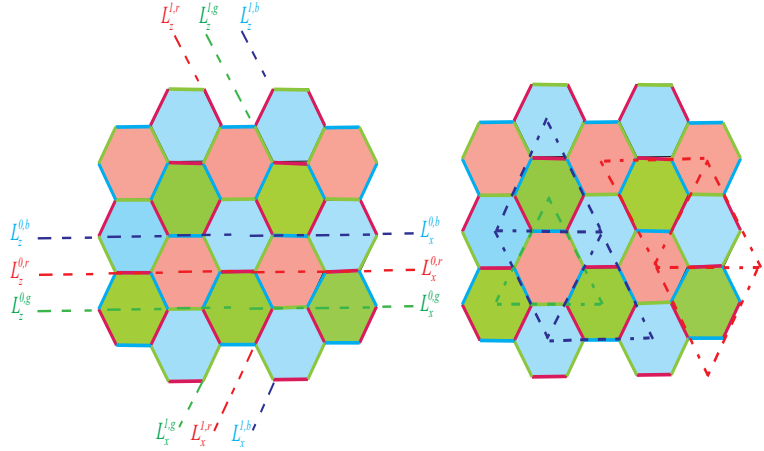}
\caption{(Color online) Left: the qubits live on the vertices of the lattice and two operators $h_x$ and $h_z$ are attached to each hexagonal plaquette. Edges of the lattice are colored by three different colors corresponding to three different colors of the hexagonal plaquettes. There are twelve non-trivial loop operators corresponding to two directions on torus and three different colors of the plaquettes.\\
Right: corresponding to each hexagonal plaquette of the lattice, there is a triangular plaquette where two qubits live on each edge. In this way, there are three triangular lattices with three different colors. Since there are two triangular plaquettes with different colors corresponding to each hexagonal plaquette, only two colors are enough for re-presentation of plaquette operators on the triangular lattices. } \label{hex}
\end{figure}

The CC model is another well-known kind of the topological lattice models with $Z_2 \times Z_2 $ topological order.  Although topological properties of this model seems similar to the TC, additional freedom degree of color in this model leads to some important differences with the TC. Specifically the CC is more efficient than the TC for computational tasks. For examples in the CC on an hexagonal lattice, all the Clifford operators can be applied in a topological way while it is not possible in the TC model \cite{bombin}.\\
The CC model can be defined on three-colorable lattices which technically called ‘colexes’ and can be generalized to arbitrary dimensions \cite{bo}. As an example we consider a hexagonal lattice where qubits live on vertices of the lattice, see figure (\ref{hex}). Corresponding to each hexagonal plaquette of the lattice which is denoted by symbol "h", we define two operators in the following form:
\begin{equation}
h_x =\prod_{i\in h} X_i~~,~~h_z =\prod_{i\in h} Z_i
\end{equation}
where $i \in h$ refers to all qubits belonging to the plaquette $h$. The Hamiltonian of the model is defined as
\begin{equation}\label{top}
H=-\sum_{h} h_x - \sum_{h}h_z .
\end{equation}
Since all plaquette operators commute with each other, the ground state of this Hamiltonian is in the following simple form:
\begin{equation}\label{c1}
|\phi_c \ra =\prod_{h}(1+h_z )|++...+\rangle
\end{equation}
where we ignore the normalization factor. Similar to the TC model, there is a loop re-presentation for the state (\ref{c1}). To this end, let us span product of operators $(1+h_z )$ in the relation (\ref{c1}) as the following form:
\begin{equation}\label{c2}
\prod_{h} (1+h_z)=1+\sum_{h}h_z +\sum_{h,h'} h_z  h'_z +... .
\end{equation}
We color all plaquettes of a hexagonal lattice by three different colors, red, blue, green so that any two neighbor plaquettes of the lattice are not in the same color, see figure (\ref{hex}, left). we also color all the edges of the hexagonal lattice with three colors so that each two plaquettes with the same color connect together with an edge with the same color. Each hexagonal plaquette of the hexagonal lattice can also be considered as a triangle plaquette of a triangular lattice where two qubits of each hexagonal plaquette live on each edge of this triangular plaquette, see figure (\ref{hex}, right).   In this way, the edges of the lattice insert in three categories corresponding to each color and we can draw a triangular lattice corresponding to each category of colored edges, see figure(\ref{hex}, right). \\
Finally, we interpret each plaquette operator of the CC model by a triangular colored loop on one of the three triangular lattices. By such a interpretation, the summation in the relation (\ref{c2}) is regarded as superposition of all possible loop operators with three different colors on the triangular lattices. A closer look to the hexagonal lattice shows that a plaquette of the initial hexagonal lattice corresponds to two different triangles of two different colored triangular lattices, see figure (\ref{hex}, right). Therefore, we have this freedom to select one of the colors for each hexagonal plaquette. By such a freedom, it is simple to show that the relation (\ref{c2}) can be interpreted as superposition of all loop operators with only two different colors. Therefore, it is well understood that the topological order in the CC is similar to the TC with additional freedom degree of color. Such a order is called $Z_2 \times Z_2$ topological order versus the $Z_2$ topological order for the TC. \\
The topological order in the CC also leads to some important properties similar to the TC such as robust degeneracy, non-local order parameter and anyonic excitations. as a sample, if we insert the hexagonal lattice on a torus, there will be many non-contractible loop operators which lead to degeneracy of the ground states. These loop operators are defined similar to the TC model with a difference that there are three kinds of  loop operators corresponding to each color, see figure (\ref{hex}, left), in the following form:
$$
L_{z}^{\sigma, r}=\prod_{i\in L^{\sigma,r}}Z_i
$$
$$
L_{z}^{\sigma, g}=\prod_{i\in L^{\sigma,g}}Z_i
$$
\begin{equation}
 L_{z}^{\sigma, b}=\prod_{i\in L^{\sigma,b}}Z_i
 \end{equation}
 where $i\in L^{\sigma, r(g,b)}$ refers to the qubits which live on a non-contractible loop around the torus on a red (green, blue)  triangular lattice in two different directions $\sigma=0~or~1$. Since three above operators are not independent so that the product of them as $L_{z}^{\sigma, r}L_{z}^{\sigma,g}L_{z}^{\sigma, b}$ is equal to a product of plaquette operators, we can generate the sixteen degenerate ground states of the model by applying only two colored loop operators as the following form:
\begin{equation}
 |\phi_{i,j,k,l}\ra=(L_{z}^{0,r})^i (L_{z}^{1,r})^j (L_{z}^{0,b})^k (L_{z}^{1,b})^l |\phi_c\rangle
 \end{equation}
where the indices $i,j,k,l=\{0,1\}$ refer to sixteen ground states of the model. Similar to TC model, there are also four non-local order parameters as the following form which can characterize different ground states of the CC model, see figure (\ref{hex}, left).
$$
L_{x}^{\sigma, r}=\prod_{i\in L^{\sigma,r}}X_i
$$
\begin{equation}
 L_{x}^{\sigma, b}=\prod_{i\in L^{\sigma,b}}X_i
 \end{equation}
 where we apply operators $X$ on non-contractible loops with two different colors.
\section{Interpolation of the CC and the TC}\label{se3}
\begin{figure}[t]
 \centering
 \includegraphics[width=10cm,height=5cm,angle=0]{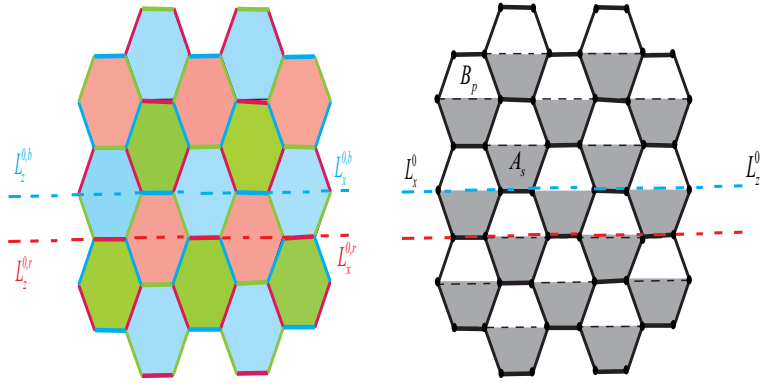}
 \caption{(Color online) Each hexagonal plaquette of the lattice has been divided to two Trapezoid-shaped parts. By a chess-pattern painting of such a lattice, a TC model can be defined where the operators $B_p$ are related to light plaquettes and the operators $A_s$ are related to dark plaquettes. Two non-contractible loops which are topologically different in the CC model convert together with product of operators $A_s$ or $B_p$ which are involved by the non-contractible loops. It show that non-contractible loops with different colors are topologically the same in the TC model. } \label{c-t}
 \end{figure}

 In this section, we consider both the TC and the CC model on the same lattice and define a new Hamiltonian as the following form:
 \begin{equation}\label{in1}
 H=-g_t H_{t}-g_c H_{c}
 \end{equation}
 where $H_t $ and $H_{c}$ are Hamiltonians of the TC and the CC model, respectively. Such a Hamiltonian is defined on the qubits which live on the vertices of a hexagonal lattice.\\ Definition of the CC on such a lattice is the same as we explained in the previous section. Since qubits live on the vertices of this lattice we can not define an ordinary TC model on such a lattice. However, there is a simple way to present a TC model on such lattice. To this end, we divide each hexagonal plaquette to two Trapezoid-shaped parts as it is shown in figure(\ref{c-t}). Then we paint all trapezoids as chess-pattern with dark and light colors. In this way, we can use a rotated version of the Kitaev model \cite{comp} where we relate an operator $B_p = Z_1 Z_2 Z_3 Z_4 $ to each light plaquette and an operator $A_s =X_1 X_2 X_3 X_4$ to each dark plaquette, see figure(\ref{c-t}). It is very simple to show that such a model is exactly the same TC model. Specifically, one can check that the ground state of this model is:
 \begin{equation}
 |\psi \ra= \prod_{p}(1+B_p)|++...+\rangle
 \end{equation}
Such a state is the same as equation (\ref{e2}) in the definition of the TC model.\\
After definition of the TC and the CC model on the same hexagonal lattice, we are ready to study the properties of the Hamiltonian (\ref{in1}). On the one hand, in limit of $g_t \gg g_c$, we have a TC model which have a four-fold degeneracy. On the other hand, in limit of $g_t \ll g_c$, we have a CC model which have a sixteen-fold degeneracy. Hence, by tuning of the coupling $g_t$ from zero to infinity we expect to see a quantum phase transition.\\
Since non-contractible loop operators generate degeneracy in the CC and the TC models, it is useful to consider how the subspace of the ground states in the CC model changes by adding the TC model. To this end, consider non-trivial loop operators in the TC and the CC on the hexagonal lattice. As it is shown in figure(\ref{c-t}), we consider four non-trivial loop operators $L_{x}^{0,r}$, $L_{z}^{0,r}$, $L_{x}^{0,b}$ and $L_{z}^{0,b}$ for the CC model and two non-trivial loop operators  $L_{x}^{0}$ and $L_{z}^{0}$ for the TC model. It is possible to describe sixteen-fold degenerate subspace of the CC by the projectors which are constructed by non-trivial loop operators as the following form:
\begin{equation}
(1+(-1)^{i}L_{x}^{0,r})(1+(-1)^{j}L_{x}^{0,b})(1+(-1)^{k}L_{z}^{0,r})(1+(-1)^{l}L_{z}^{0,b})
\end{equation}
where $i,j,k,l=\{0,1\}$. Let us compare above subspace with degenerate subspace of the TC model. As it is shown in figure (\ref{c-t}), four non-trivial operators of the CC model on the hexagonal lattice can also be considered non-trivial operators of the TC model. But there is a difference that two operators $L_{x}^{0,r}$ and $L_{x}^{0,b}$ are in the same homology class in the TC so that they convert to each other by applying product of plaquette operators $A_s$ which are involved by the two non-trivial loops. The same situation is for operators $L_{z}^{0,r}$ and $L_{z}^{0,b}$ where they convert to each other by applying product of plaquette operators $B_p$ which are involved by two non-trivial loops $L_{x}^{0,r}$ and $L_{x}^{0,b}$. In this way, by perturbing the CC model by the TC perturbation, homology classes of the non-trivial loops change so that two non-trivial loops with different colors belong to the same homology class and a quantum phase transition occurs.\\
It is also interesting if we explain equivalency of non-trivial operators in the TC model in a anyonic picture. In fact in the CC model two operators $L_{z}^{0,r}$ and $L_{z}^{0,b}$ can be interpreted as generating of two charge anyons $e_r$ and $e_b$ which turn around torus and annihilate again. In the CC model, these two charge anyons are not equivalent and they can not fuse together. Consequently they generate new degenerate states while in the TC model, there is only one charge anyon so that $e_r$ and $e_b$ are equivalent and they can fuse together.\\
 Such an interpretation of the quantum phase transition emphasizes on the role of color in the CC model so that we can claim that the CC model is topologically equivalent with two TC models with two different colors. This result is explicitly derived in \cite{fold} where authors showed that the CC model is local equivalent with two copies of the TC model. In the next section, we show this result by explicit analysis of the quantum phase transition where we show the quantum phase transition has not a topological nature. As another point, it is also useful to emphasize on the role of different gauge symmetries of the TC and the CC model in the quantum phase transition. In fact, the TC is related to the universality class of the 2D Ising model (classical) with $Z_2$ symmetry, while the CC is related to the universality class of the 2D three-body Ising model with $Z_2 \times Z_2 $ symmetry \cite{sym}. Therefore, it is expected that such a quantum phase transition can be characterized as a symmetry breaking process and has not a topological nature.\\
\section{Explicit analysis of the quantum phase transition}\label{se4}
As it was explained in the previous section, according to \cite{fold}, because of equivalency of a CC model with two copies of TC model, both models have a universal topological phase. In this section, we want to consider properties of the phase transition between the CC and the TC model to understand how the universal topological phase of them reveals in the point of quantum phase transition. We rewrite Hamiltonian Eq.(\ref{in1}) in a new basis and show that it converts to many copies of 1D
Ising model in transverse field which belongs to 2D Ising universality class. Finally, we conclude that the universal
topological phase of those models reflects in 1D nature of the phase transition.\\ 
Consider a set of quantum states as the following form:
\begin{equation}\label{s1}
|\psi_{\{r_p ,w_s\}}^{i,j}\ra=(1+(-1)^{i}L_Z ^{0})(1+(-1)^{j}L_x ^{0})\prod_{p}(1+(-1)^{r_p}B_p )\prod_{s}(1+(-1)^{w_s}A_s)|\phi_c \rangle
\end{equation}
where $|\phi_c\ra=\prod_{h}(1+h_z)|\Omega\rangle$ is the ground state of the CC model. Using $|\phi_c \ra$ in above defenition is necessary to find the effect of operators of the CC model on the state (\ref{s1}). The values of $0,1$ for indices refer to different eigenstates of this basis. Such a basis has been constructed by the projector operators related to loop operators of the TC model. It is simple to check that such states generate a complete basis. In fact, the indices $i,j$ characterize topological degenerate subspace of the TC model and $r_p$ s ($w_s$ s) refer to binary variables which correspond to each light (dark) plaquette of the lattice, we call them the virtual spins which are equivalent with anyonic excitations of the TC model. In other words, the values $0$ and $1$ for a virtual spin correspond to presence or absence of an anyon in the related plaquettes.\\
It is clear that the operators $B_p$ and $A_s$ in this new basis have a simple form. Since $B_p (1+(-1)^{r_p}B_p)=(-1)^{r_p}(1+(-1)^{r_p}B_p)$, we conclude that $B_p |\psi_{\{r_p ,w_s\}}^{i,j}\ra=(-1)^{r_p}|\psi_{\{r_p ,w_s\}}^{i,j}\ra$. We can also represent the state $|\psi_{\{r_p ,w_s\}}^{i,j}\ra$ as $|r_1 , r_2 ,...\ra|w_1 , w_2 ,...\ra$ where $|r_p\ra$ and $|w_s \ra$ are eigenstates of operator $Z$ of the virtual spins which live in the center of each light and dark plaquette of the lattice, respectively. By such a definition, Since $B_p |r_1 , r_2 ,...\ra|w_1 , w_2 ,...\ra= (-1)^{r_p}|r_1 , r_2 ,...\ra|w_1 , w_2 ,...\ra$, it is clear that the operator $B_p$ plays role of the Pauli operator $Z_p$ on a virtual spin corresponding to light plaquette $p$.\\
There is similar situation for operators $A_s$. Since $A_s(1+(-1)^{w_s}A_s)=(-1)^{w_s}(1+(-1)^{w_s}A_s)$, we will have $A_s |\psi_{\{r_p ,w_s\}}^{i,j}\ra=(-1)^{w_s}|\psi_{\{r_p ,w_s\}}^{i,j}\ra $. Consequently, the operator $A_s$ also plays role of the Pauli operator $Z_s$ on a virtual spin corresponding to dark plaquette $s$. Therefore, the TC model in the new basis is written as the following form:
\begin{equation}\label{s3}
H_{kitaev} = -\sum_{p} Z_p -\sum_{s}Z_s
\end{equation}
\begin{figure}[t]
\centering
\includegraphics[width=10cm,height=4.5cm,angle=0]{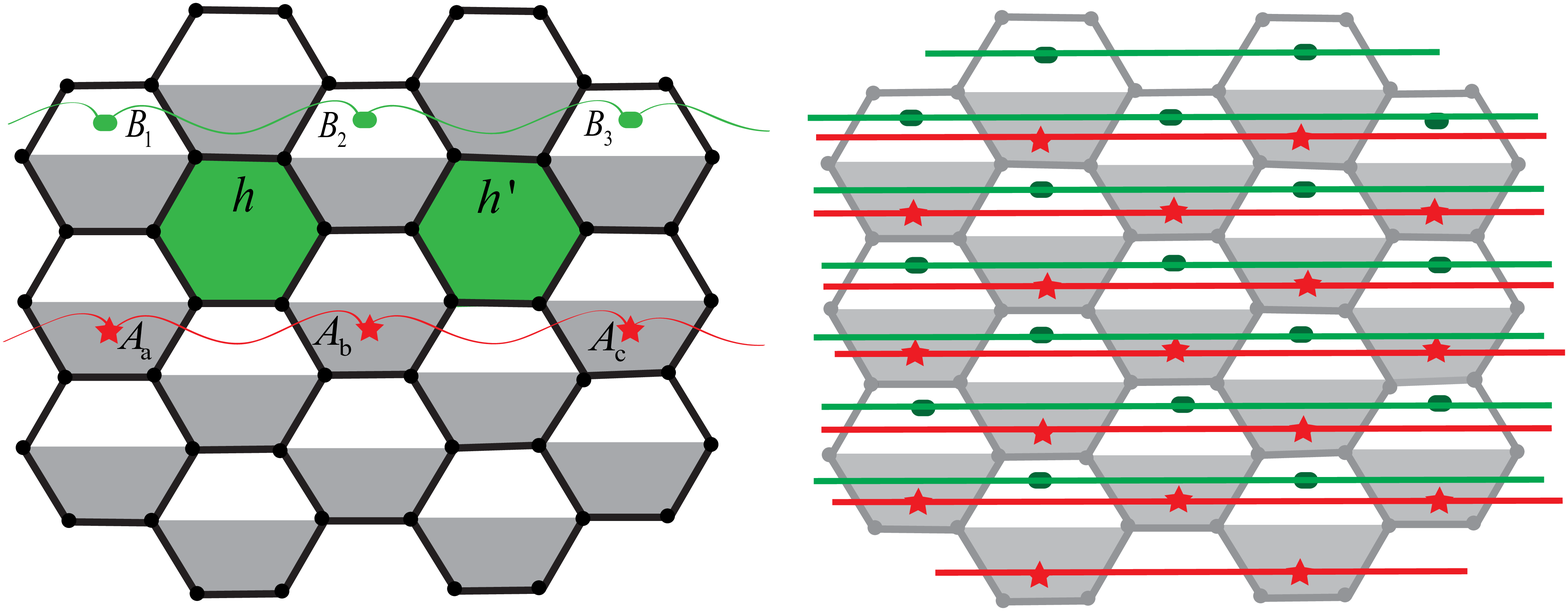}
\caption{(Color online) Left: an operator $h_x$ ($h_z$) does not commute with two light (dark) plaquette operator $B_1$ ($A_a$) and $B_2$ ($A_b$) and it is equivalent with operator $X_1 X_2$ ($X_a X_b$) on virtual spins in the center of the light (dark) plaquettes. Similarly, the operator $h'_x$ ($h'_z$) is equivalent with operator $X_2 X_3$ ($X_b X_c$) on virtual spins living in the center of light (dark) plaquettes which are denoted by green circles (red statrs). Right: By applying all hexagonal operators in the new basis, we will have $2N$ copies of 1D Ising model corresponding to each row of the lattice.} \label{Ising}
\end{figure}
It is clear that the operators of the TC model do not ever change indexes $i,j$ in the basis (\ref{s1}). In other words, we have four degenerate subspaces where there are the same forms for the Hamiltonian of the TC model. Interesting point is that there is the same situation in the Hamiltonian of the CC model. It is simple to show that the non-trivial operators $L_z^{0}$ and $L_x^{0}$ commute with all operators $h_x$ and $h_z$ in the Hamiltonian of the CC model. Consequently, the Hamiltonian of the CC model has also the same form in the four degenerate states of the TC model so that the Hamiltonian of the CC can not generate any transition between those states.\\
By attention to above argument, we consider just one of the degenerate subspaces of the TC model for rewriting Hamiltonian (\ref{in1}). Finally, we consider following states as new basis:
\begin{equation}\label{s2}
|\psi_{\{r_p\}, \{w_s\}}\ra=(1+L_Z ^{0})(1+L_x ^{0})\prod_{p}(1+(-1)^{r_p}B_p )\prod_{s}(1+(-1)^{w_s}A_s)|\phi_c \ra.
\end{equation}
  We are ready to find the new form of the CC model in above basis. In the CC model we have two operators $h_x$ and $h_z$ corresponding to each hexagon of the hexagonal lattice. we should find the effect of such operators on the states (\ref{s2}).\\
 To this end, consider a hexagon $h$ of the hexagonal lattice as it has been shown in figure(\ref{Ising}). The corresponded hexagonal operator $h_x$ involves six qubits which are common with few dark and light plaquettes of the TC model. It is clear that the operator $h_x$ commutes with all operators $A_s$ on dark plaquettes and all operators $B_p$ on light plaquettes except of two light-plaquette operators $B_1$ and $B_2$ which have only one joint qubit with the hexagon $h$, see figure (\ref{Ising}, left). Therefore, we have $h_x(1+(-1)^{r_1}B_1 )(1+(-1)^{r_2}B_2 ) =(1+(-1)^{r_1 +1}B_1 )(1+(-1)^{r_2 +1}B_2 )h_x$. Since $h_x |\phi_c \ra=|\phi_c\ra$, we conclude that the effect of operator $h_x$ on the state (\ref{s2}) leads to rising binary variables $r_1$ and $r_2$. Such a operation is the same as operator $X_1 X_2$ on the basis of virtual spins in (\ref{s2}).\\
 The situation is the same for operators $h_z$. This operator does not commute with two dark plaquettes $A_a$ and $A_b$ of the TC model, see figure (\ref{Ising}, left). Therefore, the effect of operator $h_z$ is the same as operator $X_a X_b$ on the basis (\ref{s2}). \\
 In the next step, consider another hexagonal plaquette in the neighbor of the plaquette $h$, we denote it by $h'$, see figure(\ref{Ising}, left). similar to plaquette $h$ , operator $h'_x$ dose not commute with two light-plaquette operator $B_2$ and $B_3$  so that it is equal to a operator $X_2 X_3$ on virtual spins in (\ref{s2}). If we repeat this work for other plaquettes of the lattice which are in the same row with h and h', we will have an Ising model as $\sum_{\la i,j\ra}X_i X_j$ on virtual spins which live in light-plaquettes of the corresponded row of the lattice, see figure (\ref{Ising}, right). The same situation is for operator $h'_Z$ where it is equal to operator $X_b X_c$ on virtual spins in the dark plaquettes of corresponded row of the lattice, see figure (\ref{Ising}, right).\\
 Finally, we can find the effect of operators $h_x$ and $h_z$ associating with all plaquettes of the hexagonal lattice as  $\sum_{h} (h_x + h_z) $ on the basis (\ref{s2}). As it is shown in figure(\ref{Ising}, right), the hexagonal lattice insert on a 2D torus where there are $N$ rows so that each row of the lattice contains of $N$ plaquettes. If we divide all plaquettes of the lattice into $N$ parts which correspond to rows of the lattice, we can write the Hamiltonian of the CC model in the following form:
\begin{equation}\label{s4}
H=-\sum_{i}\sum_{h\in r_i} (h_x + h_z)
\end{equation}
where $h\in r_i$ refers to plaquettes which belong to i th row of the lattice. As it is already discussed,  summation of operators $h_x$ ($h_z$) on a row of the lattice in the basis (\ref{s2}) is equal to the Ising model on virtual spins in light-plaquettes (dark-plaquettes) of that row of the lattice. Therefore , the CC model in equation (\ref{s4}) is converted to $2N$ one dimensional Ising model on dark and light virtual spins of each row of the lattice in the following form:
\begin{equation}\label{s5}
H_c = \sum_{row}(-\sum_{i\in{L_r}} X_i X_{i+1} -\sum_{j\in D_r} X_j X_{j+1})
\end{equation}
where $D_r$($L_r$) refers to dark plaquettes (light plaquettes) of the row "r".\\
After rewriting both the TC model and the CC model in the basis (\ref{s2}) according to relations (\ref{s3}) and (\ref{s5}), the Hamiltonian of the our model (\ref{in1}) is converted to the following model in the basis (\ref{s2}):
\begin{equation}\label{s6}
H = \sum_{row}\{(-\sum_{i\in{D_r}} (g_c ~X_i X_{i+1}+g_t ~Z_i) -\sum_{j\in L_r} (g_c ~X_j X_{j+1}+g_t ~Z_i)\}
\end{equation}
where there are $2N$ 1D Ising models in transverse field which have a well-known phase transition point at $g_t =g_c $ which belong to the universality class of 2D classical Ising model. \\
It is interesting to explicitly consider the ground state in two limits $g_t =0$ and $g_c =0$ according to relation(\ref{s6}). If $g_t =0$ we will have an Ising model on all rows of the lattice and the ground state will be as product of $|++...+\ra$ or $|---...-\ra$ for each row of the lattice. In the following we briefly explain how one of the ground states of TC model converts to four ground states of the CC model in effect of the quantum phase transition.\\
Consider a state $|+_a\ra=\frac{1}{\sqrt{2}}(|0_a\ra+|1_a\ra)$ on one of the virtual spins in the light or dark plaquette $a$. the state $|0_a\ra$ in the word of the basis (\ref{s2}) is equal to:
\begin{equation}\label{w1}
(1+L_Z ^{0})(1+L_x ^{0})(1+B_a)\prod_{p\neq a}(1+(-1)^{r_p}B_p )\prod_{s}(1+(-1)^{w_s}A_s)|\phi_c\rangle
\end{equation}
where $p\neq a$ refers to all plaquettes expect of the plaquette $a$. On the other hand, the state $|1_a\ra$ in the word of the basis (\ref{s2}) is equal to:
\begin{equation}\label{w2}
(1+L_Z ^{0})(1+L_x ^{0})(1-B_a)\prod_{p\neq a}(1+(-1)^{r_p}B_p )\prod_{s}(1+(-1)^{w_s}A_s)|\phi_c\rangle
\end{equation}
In this way, the state $|+_a\ra$ will be equal to:
\begin{equation}
|+_a\ra=(1+L_Z ^{0})(1+L_x ^{0})\prod_{p\neq a}(1+(-1)^{r_p}B_p )\prod_{s}(1+(-1)^{w_s}A_s)|\phi_c\rangle
\end{equation}
Therefore, it is simple to show that a state as $|++...+\ra$ on whole virtual spins in the lattice which is one of the ground states of the model is equal to:
\begin{equation}\label{w3}
|\psi_0\ra=|++...+\ra=(1+L_Z ^{0})(1+L_x ^{0})|\phi_c\rangle
\end{equation}
where it is exactly one of the ground states of the CC model. \\
Then we consider the state $|-_a\ra$. By attention to relations (\ref{w1}) and (\ref{w2}) such a state in the basis (\ref{s2}) is equal to :
\begin{equation}
|-_a\ra =B_a(1+L_Z ^{0})(1+L_x ^{0})\prod_{p\neq a}(1+(-1)^{r_p}B_p )\prod_{s}(1+(-1)^{w_s}A_s)|\phi_c\rangle
\end{equation}
This relation shows that if we change a row of the virtual spins on light plaquettes  from $|++..+\ra$ to $|--...-\ra$ it is equal to applying a row of corresponded operators $B_p$ on the state $|\psi_0\ra$ in (\ref{w3}). As it is shown in figure (\ref{c-t}), a product of operators $B_p$ on a row of the lattice is equal to product of non-trivial operators $L_{z}^{0,r}$ and  $L_{z}^{0,b}$ in the CC model. While $L_{z}^{0,r}$ and  $L_{z}^{0,b}$ are topologically equivalent in the TC model, they are different in the CC model. Therefore, one can easily show that such a operation generates another ground state of the CC model.\\ The same process occurs for the product of $A_s$ on a row of the lattice where it is equal to product of non-trivial operators $L_{x}^{0,r}$ and  $L_{x}^{0,b}$ in the CC model and generates third ground state. Finally, forth ground state of the CC model is generated by applying a product of $A_s$ and $B_p$ on a row of  dark and light plaquttes of the lattice. Therefore, one of the ground states of the TC model splits to four quantum states of the CC model in effect of the quantum phase transition.\\
In this way, we showed that the quantum phase transition between the CC and the TC model is equivalent with the phase transition in 1D Ising model in transverse field. Such a result is important in the view point of universal topological phase for the CC and the TC model. It is well-known that a 1D quantum model has not a topological order \cite{spt}. In fact for a 1D model, one can not define any topological signature such as Wilson loop or topological entanglement entropy. Consequently, our result shows that topological difference of the CC and the TC model has a 1D nature and therefore they are in the same topological phase. \\The point of phase transition at $g_t =g_c $ also shows that the CC model is strongly robust against the TC model. In fact a TC model with a small coupling $g_t < g_c$ can not lead to a quantum phase transition. While the robustness of the CC against local perturbations such as magnetic field and Ising interactions is small \cite{jah}, our result needs to a proper interpretation. It seems that since the topological phase of the TC model is the same as the CC model, the quantum phase transition in our model only relates to lifting degeneracy of the ground state. Therefore, much robustness of the CC model against the TC model relates to the much robustness of the topological degeneracy of the ground state. \\
In the next subsection we approximately study the behavior of Wilson loop in our model. we show that the Wilson loop does not behave as expected in a topological phase.
\section{The behavior of the  Wilson loop}\label{se5}
\begin{figure}[t]
\centering
\includegraphics[width=6cm,height=6cm,angle=0]{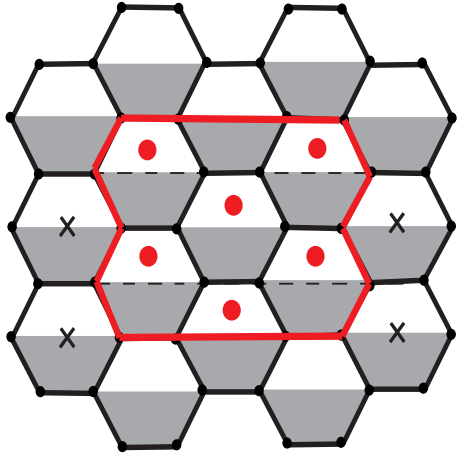}
\caption{(Color online) A Wilson loop on chess-patterned lattice is shown by red lines. The Wilson loop operator is equal to product of operators $B_p$ which is denoted by red circles. The Wilson loop operators commute with all hexagonal plaquette operators except of the operators which are near of two left and right edges of the Wilson loop and are denoted by Celtic cross. } \label{wilson}
\end{figure}

In this section, we consider behavior of the Wilson loop operators in our model. It is well-known that behavior of a Wilson loop in a perturbed topological phase is as $e^{-\alpha |\partial w_{l} |}$ where $|\partial w_l |$ is the perimeter of the Wilson loop. Such a behavior for the Wilson loop is a signature of the topological phase \cite{karimipour}.\\
We consider a Wilson loop operator in the TC model as $W_l=\prod_{i\in l}Z_i$ where $l$ is a loop on the chess-patterned hexagonal lattice, see figure (\ref{wilson}). Such a operator is equal to product of operators $B_p$ which are involved by the Wilson loop. One the one hand, it is clear that the expectation value of such an operator is equal to 1 in the ground state of the TC model. On the other hand, the expectation value of this Wilson loop is equal to zero in the ground state of the CC model. Therefore, the expectation value of the Wilson loop changes from 1 to 0 when we increase coupling of the CC model from zero to infinity. \\
Consider the case that $\gamma=\frac{g_c}{g_t}\ll 1$, the ground state of our model in such a limit can be found by a perturbation method. To this end we write a perturbed quantum state as the following form:
\begin{equation}\label{ap1}
|\psi\ra =|\phi_t \ra + \gamma \hat{O}|\phi_t\ra + \gamma^2 \hat{O}^2 |\phi_t\ra +...
\end{equation}
where the operator $\hat{O}=\sum_{h}h_x +h_z$ is the Hamiltonian of the CC model and $|\phi_t\ra$ is the normalized ground state of the pure TC model. In the next step we have to calculate the expectation value of the Wilson loop operator into the above state as $\la W_l \ra=\frac{\la \psi| W_l|\psi\ra}{\la \psi |\psi\ra} $. We approximate the ground state ((\ref{ap1}) in first order of $\gamma$ as the following form:
\begin{equation}\label{ap2}
|\psi\ra =|\phi_t \ra + \gamma \hat{O}|\phi_t\ra
\end{equation}
Since $\la \phi_t |\hat{O} |\phi_t\ra=0$, we will have:
\begin{equation}\label{ap3}
\la \psi|\psi\ra = \la \phi_t|\phi_t\ra +\gamma ^2 \la \phi_t|\hat{O}^2|\phi_t\ra=1+2P \gamma ^2
\end{equation}
where $P$ is the number of hexagonal plaquettes of the lattice. On the other hand, by the fact that $W_l |\phi_t\ra=|\phi_t \ra$ we will have:
\begin{equation}\label{ap4}
\la \psi|W_l|\psi\ra = \la \phi_t|\phi_t\ra +\gamma ^2 \la \phi_t|\hat{O}W_l \hat{O}|\phi_t\ra
\end{equation}
One can easily check that the Wilson loop operator $W_l$ commute with all hexagonal plaquettes operators $h_x$ and $h_z$ except of operators $h_x$  which cross two right and left edges of the Wilson loop, see figure (\ref{wilson}). By this and the fact that  $W_l |\phi_t\ra=|\phi_t \ra$, we conclude that:
\begin{equation}\label{ap5}
 \la \phi_t|\hat{O}W_l \hat{O}|\phi_t\ra=\la \phi_t|\hat{O_f}^2|\phi_t\ra - \la \phi_t|(\sum_{h\in E}h_x)^2|\phi_t\ra
\end{equation}
where $\hat{O_f}$ refers to the hexagonal operators which commute with the Wilson loop and $h\in E$ refers to hexagonal operators which cross two right and left edges of the Wilson loop. If we denote the length of two left and right edges of the Wilson loop by $L$, we will have:
$$
\la \phi_t|(\sum_{h\in E}h_x)^2|\phi_t\ra=L~~,
$$
\begin{equation}\label{ap5}
\la \phi_t|\hat{O_f}^2|\phi_t\ra=2P-L
\end{equation}
Finally the expectation value of the Wilson loop operators is calculated as:
$$
\la W_l \ra=\frac{\la \psi| W_l|\psi\ra}{\la \psi |\psi\ra}=(1+\gamma^2 (2P-2L)) (1+2P\gamma^2)^{-1}
$$
\begin{equation}
=1-2L\gamma^2 \approx e^{-2L \gamma^2}
\end{equation}
Since $L$ is the length of only two edges of the Wilson loop, the result shows that the expectation value of the Wilson loop does not behave as $e^{-\alpha \partial S}$ that we expect for a topological phase. In fact all Wilson loops with different perimeters but the same $L$ close to zero with a similar rate by the CC perturbation. Such a behavior shows that in effect of quantum phase transition from the TC to the CC models the topological phase does not change.
\section{Discussion}
Understanding how quantum phase transitions between different quantum models can characterize different topological classes seems a challenging task which needs to more exploration in future. In this paper, by a specific example, we considered how topological classes of different quantum phases reflect in the properties of a quantum phase transition between them.  To this end, we considered a quantum phase transition between  two well-known topological models, the CC and the TC model.  We showed that a interpolation of the CC model and the TC model is mapped to 1D quantum Ising model in a transverse field. The 1D nature of the phase transition revealed the same topological nature of the CC model and the TC model. In fact, since different quantum phases of both models was related to their different gauge symmetries, it made the possible interpolating line joining them a non-topological flow. To be more explicit, we also calculated the expectation value of the Wilson loop. The result showed that behavior of the Wilson loop captures the same topological phase of the CC and the TC models. Finally, since 1D Ising model in transverse field is a well-known model with an exact solution,  we could excactly find the point of the phase transition. The result showed that the CC model is strongly robust against the TC model.

 \end{document}